\documentclass[12pt,a4paper]{article}
\usepackage{graphicx}
\usepackage{amsmath}

\input{tcilatex}

\begin{document}

\title{Larmor radius effects on impurity transport in turbulent plasmas}
\author{{\small M. Vlad and F. Spineanu } \\
{\small National Institute for Laser, Plasma and Radiation Physics, }\\
{\small Association Euratom-MEC, P.O.Box MG-36, Magurele, Bucharest, Romania}}
\date{}
\maketitle

\begin{abstract}
Test particle transport determined by the Lorentz force in turbulent
magnetized plasmas is studied. The time dependent diffusion coefficient,
valid for the whole range of parameters, is obtained by developing the
decorrelation trajectory method. The effects of Larmor gyration in the
presence of trapping are analyzed and several transport regimes are
evidenced. \vspace{0.20in}



\vspace*{0.30in}

\end{abstract}

\section{Introduction}

Impurity control in magnetically confined plasmas is a very important issue
for the development of fusion reactors. Impurity behaviour is a complex
problem related to confinement and transport of the bulk ions and electrons
in plasma and to plasma-wall interaction. A very strong experimental effort
(see e.g. \cite{imp1}-\cite{imp9}) lead to the conclusion that this process
is far from being understood on the basis of the existing theoretical
models. In particular, the experimental results for the diffusion
coefficients are much larger than the neoclassical prediction, especially in
the edge plasma, showing the presence of anomalous transport.

We analyze here the particular topic of impurity transport in turbulent
plasmas using test particle approach. Particle motion in a stochastic
potential was extensively studied in the guiding center approximation \cite
{mccomb}-\cite{K02}. It is well known since many years \cite{kraichnan}
that, for slowly varying or large amplitude turbulence, the $\mathbf{E}%
\times \mathbf{B}$ drift determines a process of dynamical trapping of the
trajectories. It consists of trajectory winding around the extrema of the
stochastic potential and strongly influences the transport. Important
progresses in the study of this nonlinear process were recently obtained.
New statistical methods were developed \cite{V98}, \cite{VS03} that
permitted to determine the asymptotic diffusion coefficient and also the
correlation of the Lagrangian velocity and the time dependent (running)
diffusion coefficient. It was shown that the trapping process determines the
decrease of the diffusion coefficient and the change of its scaling in the
parameters of the stochastic field. These methods were extended to more
complex models, which consider, besides the $\mathbf{E}\times \mathbf{B}$
drift, particle collisions \cite{V00}, an average velocity \cite{V01} or the
parallel motion \cite{VNF}, and also to the study of the collisional
particle diffusion in stochastic magnetic fields \cite{V03}. The conclusion
of these studies is that the trapping combined with a decorrelation
mechanism determines anomalous transport regimes. In these regimes the
dependence of the diffusion coefficient $D$ on the parameters describing the
decorrelation is\emph{\ inverted }due to trapping in the sense that a
decrease of $D$ \ appearing in the quasilinear regime is transformed into an
\ ''anomalous'' increase of $D$ in the nonlinear regime \cite{VSPS}.

All these studies are based on the guiding center approximation for particle
motion that considers the Larmor radius negligible. This approximation is
not adequate for the impurity ions which can have Larmor radii comparable or
larger than the correlation length of the turbulence and cyclotron periods
comparable with the turnover time of the $\mathbf{E}\times \mathbf{B}$
motion. In these conditions the trajectories have to be determined from the
Lorentz force. The aim of this paper is to determine the effects of finite
Larmor radius on particle transport in a turbulent magnetized plasmas. In
particular, we analyze the influence on the trapping process and compare the
characteristics of the transport induced by the Lorentz force (\emph{Lorentz
transport}\textit{)} with those obtained in the guiding center approximation
(\emph{drift transport}\textit{).}\ The time dependent diffusion coefficient
is obtained as a function of the turbulence and ion's parameters by
extending the decorrelation trajectory method developed for the drift
transport \cite{V98}, \cite{PPCF04}. The transport regimes for a large range
of parameters are determined.

The paper is organized as follows. Section 2 contains the basic equations,
the statistical approach and the derivation of the general expression for
the time dependent diffusion coefficient. The results are presented in
Sections 3-6. First, in Section 3, a static potential is considered. We show
that the trapping exists even at large values of the Larmor radius and that
it determines a subdiffusive transport, as in the drift approximation, but
with a time dependence of the diffusion coefficient strongly influenced by
the gyration motion. Then we examine the dependence of the asymptotic
diffusion coefficient on the three dimensionless parameters (see Section 2
for their definitions) that characterize this process and show that several
transport regimes appear. The dependence on the first parameter (the Kubo
number) that describes the effect of the time dependence of the stochastic
potential is analyzed in Section 4 where two regimes are evidenced. The
dependence of $D$ on the initial kinetic energy of the ions (the second
parameter) is presented in Section 5 and in Section 6 the effect of the
normalized cyclotron frequency that essentially describes the specific
charge of the ions is examined. The conclusions are summarized in Section 7.

\section{Basic equations and statistical approach}

We consider a constant confining magnetic field directed along $z$ axis, $%
\mathbf{B}=B\mathbf{e}_{z}$ (slab geometry) and an electrostatic turbulence
represented by an electrostatic potential $\ \phi ^{,}(\mathbf{x},t),$ where 
$\mathbf{x}\equiv (x_{1},x_{2})$ are the Cartesian coordinates in the plane
perpendicular to $\mathbf{B}.$ The motion of an ion with charge $q$ and mass 
$m$ is determined by the Lorentz force: 
\begin{equation}
m\frac{d^{2}\mathbf{x}(t)}{dt^{2}}=q\left\{ -\mathbf{\nabla }\phi ^{,}(%
\mathbf{x},t)+\mathbf{u\times B}\right\}  \label{1}
\end{equation}
where $\mathbf{x}(t)$ is the ion trajectory, $\mathbf{u}(t)=d\mathbf{x}%
(t)/dt $ is its velocity and $\mathbf{\nabla }$ is the gradient in the $%
(x_{1},x_{2})$ plane. The initial conditions are 
\begin{equation}
\mathbf{x}(0)=\mathbf{0,\quad u}(0)=\mathbf{u}_{0}.  \label{ic}
\end{equation}
This equation is transformed into a system of first order equations for the
position and the velocity of the ion 
\begin{equation}
\frac{du_{i}}{dt}=-\frac{q}{m}\frac{\partial \phi ^{,}(\mathbf{x},t)}{%
\partial x_{i}}+\Omega \varepsilon _{ij}u_{j}  \label{1v}
\end{equation}
\begin{equation}
\frac{dx_{i}}{dt}=u_{i}  \label{1x}
\end{equation}
where $\Omega =qB/m$ is the cyclotron frequency and $\varepsilon _{in}$ is
the antisymmetric tensor ($\varepsilon _{12}=-\varepsilon _{21}=1,$ $%
\varepsilon _{11}=\varepsilon _{22}=0).$ Introducing the instantaneous
Larmor radius defined by 
\begin{equation}
\rho _{i}(t)\equiv -\varepsilon _{ij}\frac{u_{j}(t)}{\Omega },  \label{ro}
\end{equation}
the guiding center position $\mathbf{\xi }(t)\equiv \mathbf{x}(t)-\mathbf{%
\rho }(t)$ and $\phi (\mathbf{x},t)\equiv \phi ^{,}(\mathbf{x},t)/B,$ \ the
system becomes 
\begin{equation}
\frac{d\xi _{i}}{dt}=-\varepsilon _{ij}\frac{\partial \phi (\mathbf{\xi }+%
\mathbf{\rho ,}t)}{\partial \xi _{j}}  \label{1csi}
\end{equation}
\begin{equation}
\frac{d\rho _{i}}{dt}=\varepsilon _{ij}\left[ \frac{\partial \phi (\mathbf{%
\xi }+\mathbf{\rho }.t)}{\partial \xi _{j}}+\Omega \rho _{j}\right] .
\label{1ro}
\end{equation}

The electrostatic potential $\phi (\mathbf{x},t)$ is a stochastic field and
thus Eqs. (\ref{1csi}-\ref{1ro}) are Langevin equations. The solution
consists, in principle, in determining the statistical properties of the
ensembles of trajectories, each one obtained by integrating Eqs. (\ref{1csi}-%
\ref{1ro}) for a realization of the stochastic potential. We will determine
here the mean square displacement and the time dependent diffusion
coefficient for the guiding center trajectories $\mathbf{\xi }(t).$\ These
statistical quantities can also be determined for particle trajectories $%
\mathbf{x}(t)$ and for the Larmor radius $\mathbf{\rho }(t)$ but they are
not physically relevant.

The potential is considered to be a stationary and homogeneous Gaussian
stochastic field, with zero average. Such a stochastic field is completely
determined by the two-point Eulerian correlation function, $E(\mathbf{x},t),$
defined by 
\begin{equation}
E(\mathbf{x},t)\equiv \left\langle \phi (\mathbf{x}^{,},t^{,})\,\phi (%
\mathbf{x}^{,}+\mathbf{x},t^{,}+t)\right\rangle .  \label{pec}
\end{equation}
The average $\left\langle ...\right\rangle $ is the statistical average over
the realizations of $\phi (\mathbf{x},t).$ The statistical properties of the
drift velocity components 
\begin{equation}
v_{i}^{dr}(\mathbf{x,}t)\equiv -\varepsilon _{ij}\frac{\partial \phi (%
\mathbf{x,}t)}{\partial x_{j}}  \label{vdr}
\end{equation}
are completely determined by those of the potential; they are stationary and
homogeneous Gaussian stochastic fields like $\phi (\mathbf{x},t)$. The
two-point Eulerian correlations of the drift velocity components, $E_{ij}(%
\mathbf{x},t)\equiv \left\langle v_{i}^{dr}(\mathbf{x}^{,},t^{,})%
\,v_{j}^{dr}(\mathbf{x}^{,}+\mathbf{x},t^{,}+t)\right\rangle ,$ and \ the
potential-velocity correlations, $E_{\phi i}(\mathbf{x},t)\equiv
\left\langle \phi (\mathbf{x}^{,},t^{,})\,\,v_{i}^{dr}(\mathbf{x}^{,}+%
\mathbf{x},t^{,}+t)\right\rangle ,$\emph{\ } are obtained using Eq. (\ref
{vdr}) as: 
\begin{eqnarray}
E_{ij}(\mathbf{x},t) &=&-\varepsilon _{in}\varepsilon _{jm}\frac{\partial
^{2}E(\mathbf{x},t)}{\partial x_{n}\partial x_{m}},  \label{ecvel} \\
E_{\phi i}(\mathbf{x},t) &=&-E_{i\phi }(\mathbf{x},t)=-\varepsilon _{in}%
\frac{\partial E(\mathbf{x},t)}{\partial x_{n}}.  \notag
\end{eqnarray}
The Eulerian correlation of the drift velocity (\ref{ecvel}) evidences three
parameters: the amplitude $V=\sqrt{E_{11}(\mathbf{0},0)}$, the correlation
time $\tau _{c},$ which is the decay time of the Eulerian correlation, and
the correlation length $\lambda _{c},$ which is the characteristic decay
distance. These parameters combine in a dimensionless Kubo number 
\begin{equation}
K=\frac{V\tau _{c}}{\lambda _{c}}\,=\frac{\tau _{c}}{\tau _{fl}}  \label{K}
\end{equation}
which is the ratio of $\tau _{c}$ to the average time of flight of the
particles ($\tau _{fl}=\lambda _{c}/V$) over the correlation length. Using
these parameters of the stochastic field, Eqs.(\ref{1csi}-\ref{1ro}) are
written is dimensionless form as 
\begin{equation}
\frac{d\xi _{i}}{dt}=-\varepsilon _{ij}\frac{\partial \phi (\mathbf{\xi }+%
\mathbf{\rho ,}t)}{\partial \xi _{j}}  \label{1csin}
\end{equation}
\begin{equation}
\frac{d\rho _{i}}{dt}=\varepsilon _{ij}\left[ \frac{\partial \phi (\mathbf{%
\xi }+\mathbf{\rho }.t)}{\partial \xi _{j}}+\overline{\Omega }\rho _{j}%
\right]   \label{1ron}
\end{equation}
where the normalization parameters are $\tau _{fl}$ for time, $\lambda _{c}$
for distances, $V$ for the drift velocity and 
\begin{equation}
\overline{\Omega }=\Omega \tau _{fl}.  \label{obar}
\end{equation}
The same notations are kept for the normalized quantities.

We note that the equation for the guiding center trajectory (\ref{1csin}) is
similar with that obtained in the guiding center approximation, with the
difference that the argument of the potential is the particle trajectory $%
\mathbf{x}(t)\equiv \mathbf{\xi }(t)\mathbf{+\rho }(t)$ instead of $\mathbf{%
\xi }(t).$ The equation for the Larmor radius (\ref{1ron}) describes a
cyclotron motion that has the radius and the frequency dependent of the
stochastic potential. In the 2-dimensional case studied here, both equations
are of Hamiltonian type (the two components of $\mathbf{\xi }(t)$ are
conjugated variables as well as the two components of the Larmor radius $%
\mathbf{\rho }(t)$%
\begin{equation}
\frac{d\xi _{i}}{dt}=-\varepsilon _{ij}\frac{\partial H(\mathbf{\xi },%
\mathbf{\rho })}{\partial \xi _{j}},\quad \frac{d\rho _{i}}{dt}=\varepsilon
_{ij}\frac{\partial H(\mathbf{\xi },\mathbf{\rho })}{\partial \rho _{j}}.
\label{sysh}
\end{equation}
They have the same Hamiltonian function 
\begin{equation}
H(\mathbf{\xi },\mathbf{\rho })=\phi (\mathbf{\xi }+\mathbf{\rho })+\frac{%
\overline{\Omega }}{2}(\rho _{1}^{2}+\rho _{2}^{2}),  \label{has}
\end{equation}
which is the energy of the particle. The Hamiltonian depends on $\mathbf{\xi 
}$ and $\mathbf{\rho }$\ and thus the two Hamiltonian systems (\ref{1csin})
and (\ref{1ron}) are coupled. For each system the other variable introduces
a time dependence in $H(\mathbf{\xi },\mathbf{\rho })$ which perturbs the
regular motion that is obtained in the absence of interaction. The
perturbation can be very strong leading to a chaotic motion of the guiding
centers.

Particle motion is thus determined by three dimensionless parameters: $K,$ $%
\overline{\rho }$ and $\overline{\Omega }.$\ The first one, the Kubo number $%
K,$ does not appear in the equations, but only in the statistical
description of the stochastic potential. It describes the effect of time
variation of the stochastic potential. The second parameter $\overline{\rho }
$ is the initial Larmor radius normalized with the correlation length and
appears in the initial condition (\ref{ic}), which is written as 
\begin{equation}
\rho _{1}(0)=\overline{\rho }\cos (\alpha ),\quad \rho _{2}(0)=\overline{%
\rho }\sin (\alpha ),\quad \mathbf{\xi }(0)=-\mathbf{\rho }(0)  \label{icn}
\end{equation}
where $\overline{\rho }=\frac{\left| \mathbf{\rho }(0)\right| }{\lambda _{c}}%
=\left| \mathbf{u}_{0}\right| /V\overline{\Omega }$ \ and $\alpha $
determines the orientation of the initial velocity (the angle between $%
\mathbf{u}_{0}$ and the $x_{1}$ axis is $\pi /2-\alpha $). $\overline{\rho }$
is related to the initial kinetic energy of the particles. The third
parameter $\overline{\Omega }$ defined in Eq.(\ref{obar}) is the cyclotron
frequency normalized with $\tau _{fl}$ and describes the relative importance
of the cyclotron and drift motion (second and respectively first term in \ref
{1ron})) in the evolution of the Larmor radius. 

Starting from the statistical description of the stochastic potential, we
will determine the correlation of the Lagrangian drift velocity,\ defined by:

\begin{equation}
L_{ij}(t)\equiv \left\langle v_{i}^{dr}\left[ \mathbf{x}(0),0\right]
)v_{j}^{dr}\left[ \mathbf{x}(t),t\right] \right\rangle .  \label{CL}
\end{equation}
The mean square displacement of the guiding center and its time dependent
diffusion coefficient are integrals of this function: 
\begin{equation}
\left\langle \xi _{i}^{2}(t)\right\rangle =2\int_{0}^{t}d\tau \;L_{ii}(\tau
)\;(t-\tau ),  \label{MSD}
\end{equation}
\begin{equation}
D_{i}(t)=\int_{0}^{t}d\tau \;L_{ii}(\tau ),  \label{D}
\end{equation}
provided that the process is stationary \cite{Taylor}.

The guiding center approximation obtained by taking $\mathbf{\rho =0}$ in
Eq.(\ref{1csin}) was recently studied by developing a semi-analytical
approach, the decorrelation trajectory method, \cite{V98}, \cite{PPCF04}.
Using this approach an important progress was obtained in the understanding
of the intrinsic trapping process specific to the $\mathbf{E\times B}$
drift. We present here a generalization of the decorrelation trajectory
method that applies to the Lorentz transport described by Eqs. (\ref{1csin}-%
\ref{1ron}).

The Langevin equations (\ref{1csin}-\ref{1ron}) for given values of the
parameters $\overline{\Omega },$ $\overline{\rho }$ and $K$ is studied in
subensembles (S) of realizations of the stochastic field, which are
determined by given values of the potential and of the drift velocity in the
starting point of the trajectories: 
\begin{equation}
\phi (\mathbf{0},0)=\phi ^{0},\quad \mathbf{v}^{dr}(\mathbf{0},0)=\mathbf{v}%
^{0}.  \label{2}
\end{equation}
All the trajectories contained in a subensemble have the same initial
energy. The stochastic (Eulerian) potential and drift velocity in a
subensemble (S) defined by condition (\ref{2}) are Gaussian fields but
non-stationary and non-homogeneous, with space and time dependent averages.
These averages depend on the parameters of the subensemble and are
determined by the Eulerian correlation of the potential 
\begin{equation}
\Phi (\mathbf{x,}t;S)\equiv \left\langle \phi (\mathbf{x},t)\right\rangle
_{S}=\phi ^{0}\frac{E(\mathbf{x},t)}{E(\mathbf{0},0)}+v_{1}^{0}\frac{%
E_{1\phi }(\mathbf{x},t)}{E_{11}(\mathbf{0},0)}+v_{2}^{0}\frac{E_{2\phi }(%
\mathbf{x},t)}{E_{22}(\mathbf{0},0)},  \label{fims}
\end{equation}
\begin{equation}
V_{i}(\mathbf{x,}t;S)\equiv \left\langle v_{i}^{dr}\left[ \mathbf{x},t\right]
\right\rangle _{S}=-\varepsilon _{ij}\frac{\partial \Phi (\mathbf{x,}t;S)}{%
\partial x_{j}}  \label{vms}
\end{equation}
where $\left\langle ...\right\rangle _{S}$ is the statistical average aver
the realizations that belong to (S). They are equal to the corresponding
imposed condition (\ref{2}) in $\mathbf{x=0}$ and $t=0$ and decay to zero at
large distance and/or time. The existence of an average Eulerian drift
velocity in the subensemble determines an average motion, i.e. an average
Lagrangian drift velocity $\mathbf{V}^{L}(t;S)\equiv \left\langle \mathbf{v}%
^{dr}\left[ \mathbf{x}(t),t\right] \right\rangle _{S}$ . The correlation of
the Lagrangian drift  velocity for the whole ensemble of realizations (\ref
{CL}) can be written as 
\begin{equation}
L_{ij}(t)=\int \int d\phi ^{0}\,d\mathbf{v}^{0}\,P_{1}(\phi ^{0},\mathbf{v}%
^{0};\mathbf{0},0)\;v_{i}^{0}V_{j}^{L}(t;S)\,  \label{3}
\end{equation}
where $P_{1}(\phi ^{0},\mathbf{v}^{0};\mathbf{0},0)\,$ is the probability
that a realization belongs to the subensemble (S). The average Lagrangian
drift velocity $\mathbf{V}^{L}(t;S)$ is determined using an approximation
that essentially consists in neglecting the fluctuations of the trajectories
in (S). This approximation is validated in \cite{VS03} where the
fluctuations of the trajectories in (S) are taken into account in a more
complicated and precise method. It is shown that they lead to a weak
modification of the diffusion coefficients $D(t),$ although they strongly
change $\mathbf{V}^{L}(t;S)$. Introducing the average guiding center
trajectory in (S), $\mathbf{\Xi (}t;S)\equiv \left\langle \mathbf{\xi }%
(t)\right\rangle _{S},$ and the average Larmor radius in (S), $\mathbf{\Pi }%
(t;S)\equiv \left\langle \mathbf{\rho }(t)\right\rangle _{S},$ the equations
of motion can be averaged over the realizations in (S) in this approximation
and yield 
\begin{equation}
\frac{d\Xi _{i}}{dt}=-\varepsilon _{ij}\frac{\partial \Phi (\mathbf{\Xi }+%
\mathbf{\Pi ,}t;S)}{\partial \Xi _{j}},  \label{1csiS}
\end{equation}
\begin{equation}
\frac{d\Pi _{i}}{dt}=\varepsilon _{ij}\left[ \frac{\partial \Phi (\mathbf{%
\Xi }+\mathbf{\Pi ,}t;S)}{\partial \Xi _{j}}+\overline{\Omega }\Pi _{j}%
\right] .  \label{1roS}
\end{equation}
The initial conditions for the two components of the subensemble average
trajectory are obtained from Eq. (\ref{icn}) 
\begin{equation}
\Pi _{1}(0)=\overline{\rho }\cos (\alpha ),\quad \Pi _{2}(0)=\overline{\rho }%
\sin (\alpha ),\quad \mathbf{\Xi }(0)=-\mathbf{\Pi }(0).  \label{icS}
\end{equation}
Since the orientation of the initial velocity $\mathbf{u}_{0}$ is arbitrary,
we will consider in each realization in (S) initial conditions with uniform
distribution of $\alpha $ over the interval $[0,2\pi )$.

This approximation ensures the conservation of the subensemble average
energy of the particles.

Considering for simplicity an isotropic stochastic potential with the
Eulerian correlation depending $\left| \mathbf{x}\right| ,$ a diagonal
correlation tensor is obtained for the Lagrangian drift velocity, $%
L_{ij}(t)=\delta _{ij}L(t),$ and the following expressions for the time
dependent (running) diffusive coefficient: 
\begin{equation}
D(t)\equiv D_{B}F(t),  \label{ddt}
\end{equation}
\begin{equation}
L(t)=V^{2}\frac{dF(t)}{dt}  \label{L}
\end{equation}
where 
\begin{equation}
F(t)=\frac{1}{2\left( 2\pi \right) ^{3/2}}\int_{-\infty }^{\infty }\!d\phi
^{0}\exp \left( -\frac{\left( \phi ^{0}\right) ^{2}}{2}\right)
\int_{0}^{\infty }\!dv\,v^{2}\exp \left( -\frac{v^{2}}{2}\right)
\int_{0}^{2\pi }d\alpha \,\Xi _{1}(t;S)  \label{fdt}
\end{equation}
and $\Xi _{1}(t;S)$ is the component of the solution of Eqs.(\ref{1csiS}-\ref
{1roS}) along the initial average drift velocity $\mathbf{v}^{0}$ and $%
v=\left| \mathbf{v}^{0}\right| .$

We have thus determined the correlation of the Lagrangian drift velocity
(for ions with mass $m$, charge $q$ and given initial kinetic energy)
corresponding to given Eulerian correlation $E(\mathbf{x},t)$ of the
stochastic potential. Explicit results for $L(t)$ and $D(t)$ are obtained by
effectively calculating the average trajectories in (S), solutions of Eqs. (%
\ref{1csiS}-\ref{icS}), and the weighted average (\ref{fdt}). This procedure
appears to be very similar with a direct numerical study of the simulated
trajectories. There are however essential differences. The average
trajectories are obtained for a rather smooth and simple potential and the
number of trajectories is much smaller than in the numerical study due to
the weighting factor determined analytically. This reduced very much the
calculation time, such that it can be performed on PC. A compuder code is
developed for explicit calculation of $D(t)$ for given values of the
parameters $K,$ $\overline{\rho },$ $\overline{\Omega }$ and prescribed
Eulerian correlation of the potential.

The results presented in next sections are for $E(\mathbf{x},t)=1/(1+\mathbf{%
x}^{2}/2)\exp (-t/K).$ As shown in \cite{PPCF04} the shape of the Eulerian
correlation of the potential determines the strength of the trapping
represented by the exponent of the time decay of the diffusion coefficient
in the static case. However the general behavior of the decorrelation
trajectories and of $L(t)$ and $D(t)$ are the same for all correlations.

\section{Transport by Lorentz force}

We consider here a static potential ($K,\tau _{c}\rightarrow \infty )$ and
compare the results with those obtained in the guiding center approximation
(drift transport) \cite{V98}, \cite{PPCF04}. The aim is to identify the
effects of the Larmor radius.

In the frame of the decorrelation trajectory method, the difference between
the Lorentz and the drift transport consists in the equations for the
decorrelation trajectories. The trajectories obtained from Eqs. (\ref{1csiS}-%
\ref{1roS}) are much complicated than in the guiding center approximation.
The important simplification introduced in the latter case that actually
reduces the number of parameters at one and eliminates the time dependence
of the stochastic potential (see \cite{V98}, \cite{PPCF04}) cannot be
applied in the present case. The trajectories effectively depend on the six
parameters $\alpha ,$ $u,$ $\phi ^{0},$ $\overline{\Omega },$ $\overline{%
\rho }$ and $K.$ In the static case there are closed periodic trajectories
of the guiding center $\mathbf{\xi }(t),$ even at large Larmor radii which
shows that trapping exists. The initial drift velocity does not influence
only the period but also the size and the shape of the paths. The
orientation of the initial velocity of the particle, $\alpha ,$ strongly
influences the trajectories obtained with the initial condition (\ref{icS})
because the average produced by the gyration is different for different
values of $\alpha .$

\vspace*{0.10in}

\begin{center}
\resizebox{3.2in}{!}{\includegraphics{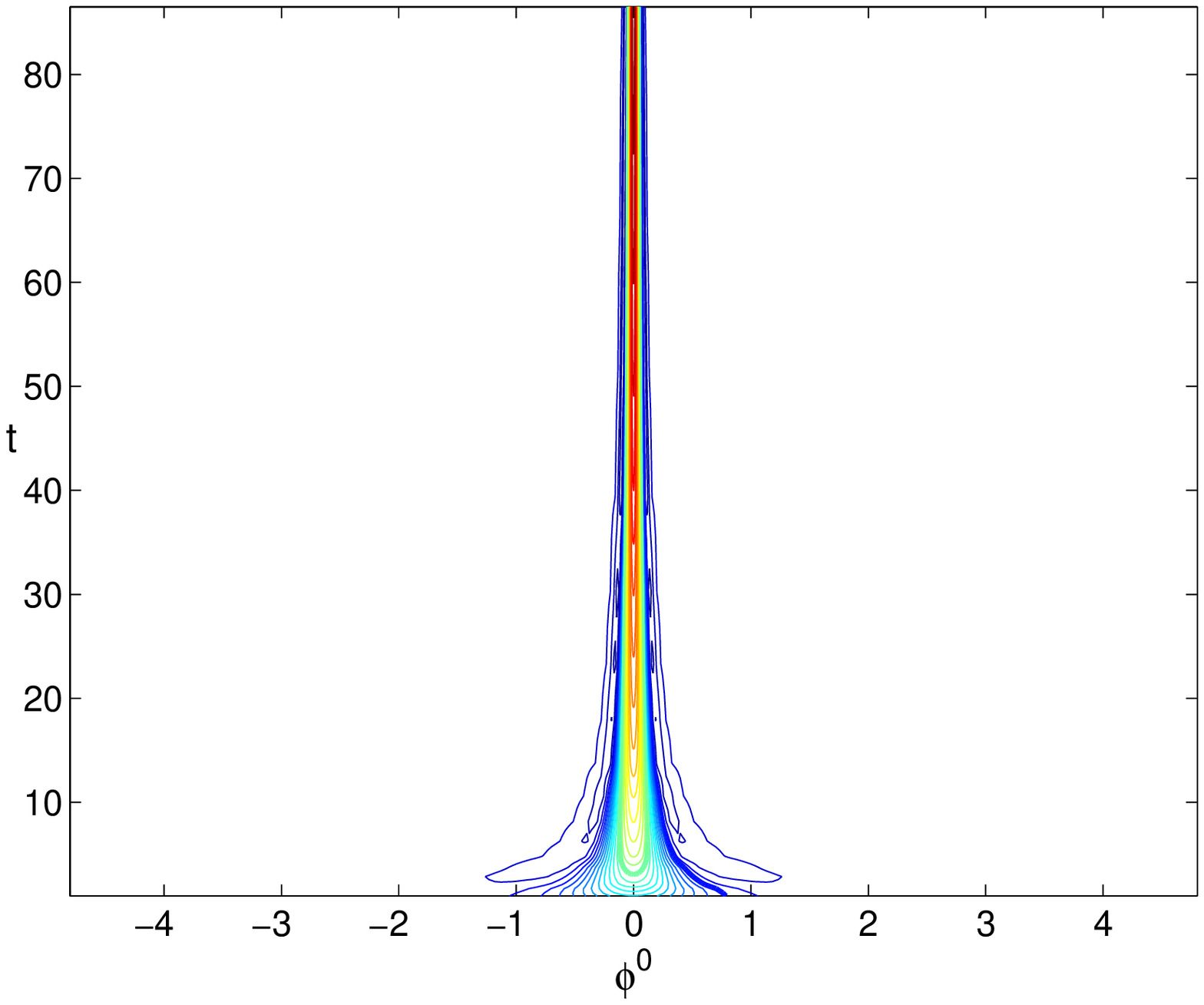}}


a
\end{center}


\begin{center}
\resizebox{3.2in}{!}{\includegraphics{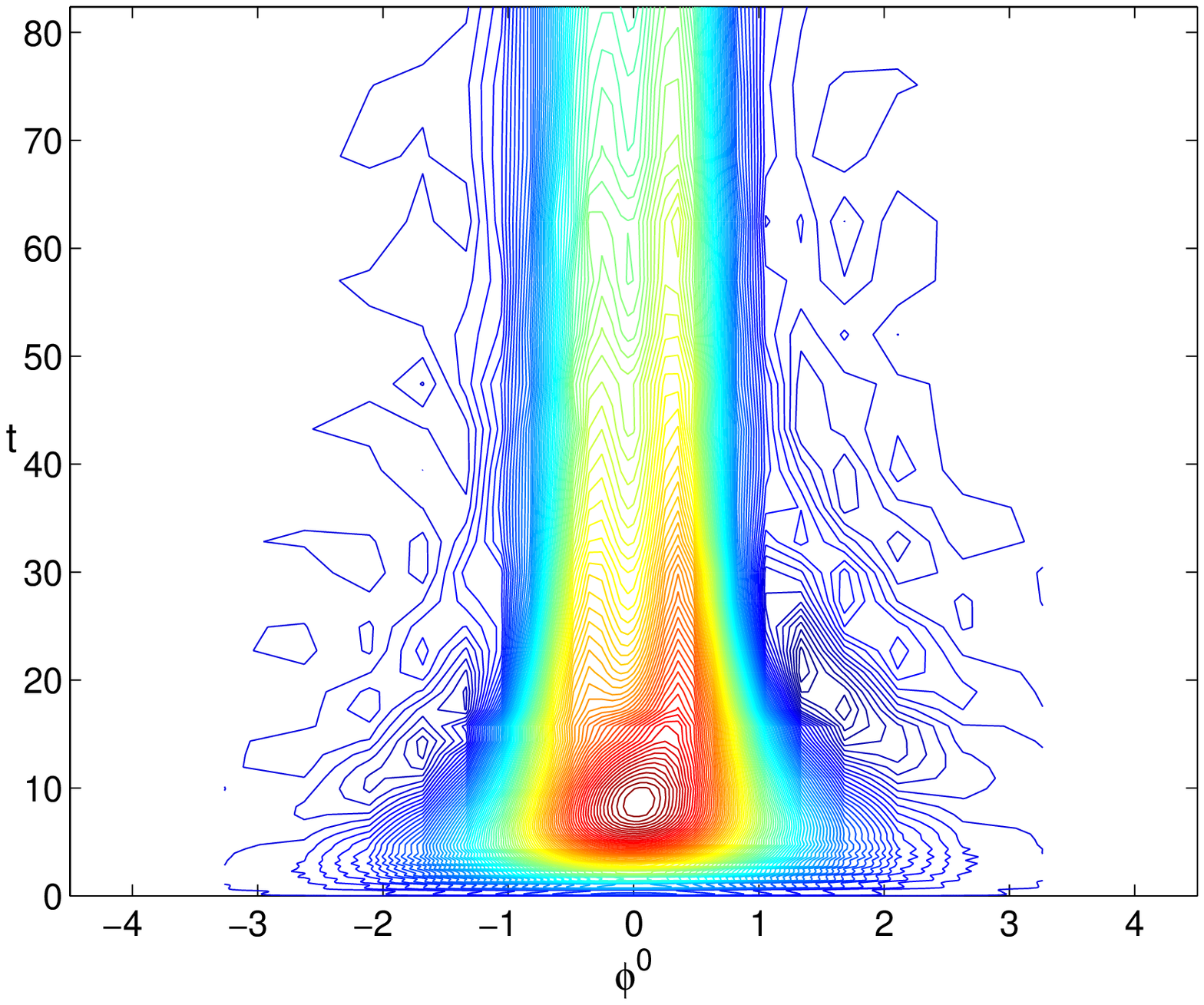}}


b
\end{center}


\begin{center}
Figure 1: The contour plot of the function $f(\phi ^{0},t)$ for the drift
(a) and Lorentz (b) transport.
\end{center}

\vspace*{0.1in}

The function $F(t)$ in Eq. (\ref{fdt}) is obtained by summing the
contribution, $f(\phi ^{0},t),$ of all the decorrelation trajectories that
start from a point where the potential is $\phi ^{0}.$ This function $f(\phi
^{0},t)$ gives details of the diffusion process showing the trapping. The
contour plot of this function is represented in Fig. 1.a. for the drift
transport and in Fig. 1.b. for the Lorentz transport. One can see that in
the first case the contributions of the large values of $\left| \phi
^{0}\right| $ are eliminated progressively due to trapping and $f(\phi
^{0},t)$ shrinks continuously as time increase such that thinner and thinner
intervals of $\phi ^{0}$ centered around zero contribute to $D(t)$. The
Lorentz transport is characterized by a completely different pattern for
this function (Fig. 1.b). At small time it has a Gaussian shape that
increase up to a maximum. This increase is not smooth but performed in
steps. At later times, $f$ becomes unsymmetrical and has a minimum at $\phi
^{0}=0.$ It decays continuously in time maintaining a large range of $\phi
^{0}$. Thus, the trapping process is completely different for the Lorentz
transport. At each value of $\phi ^{0}$ in a large interval, the cyclotron
motion determines large, open trajectories as well as small, closed ones,
depending on the values of $\alpha $ and $v.$ The contributions of small
trajectories are progressively eliminated by mixing determining the decay of
the function $f(\phi ^{0},t).$

The time dependent diffusion coefficient is presented in Fig. 2 for the
Lorentz transport (continuous line) and for the drift transport (dashed
line). The time dependence of the Lorentz diffusion coefficient is rather
complex and a strong influence of the Larmor radius can be observed. \ At
small time the diffusion coefficient increases nonuniformly, in steps.
Averaging these steps a linear time dependence can be observed, similar with
that obtained in the drift transport. This behavior extends to times much
longer than the flight time. The maximum of $D(t)$ is at about $7\tau
_{fl}.\,\ $At later times a decay of $D(t)$ appears with a time dependence
that is approximately the same as in the drift case, but with $D(t)$ larger
with a factor of about $2.$ Thus the transport in the static case is
subdiffusive.

\vspace*{0.1in}

\begin{center}
\resizebox{3.8in}{!}{\includegraphics{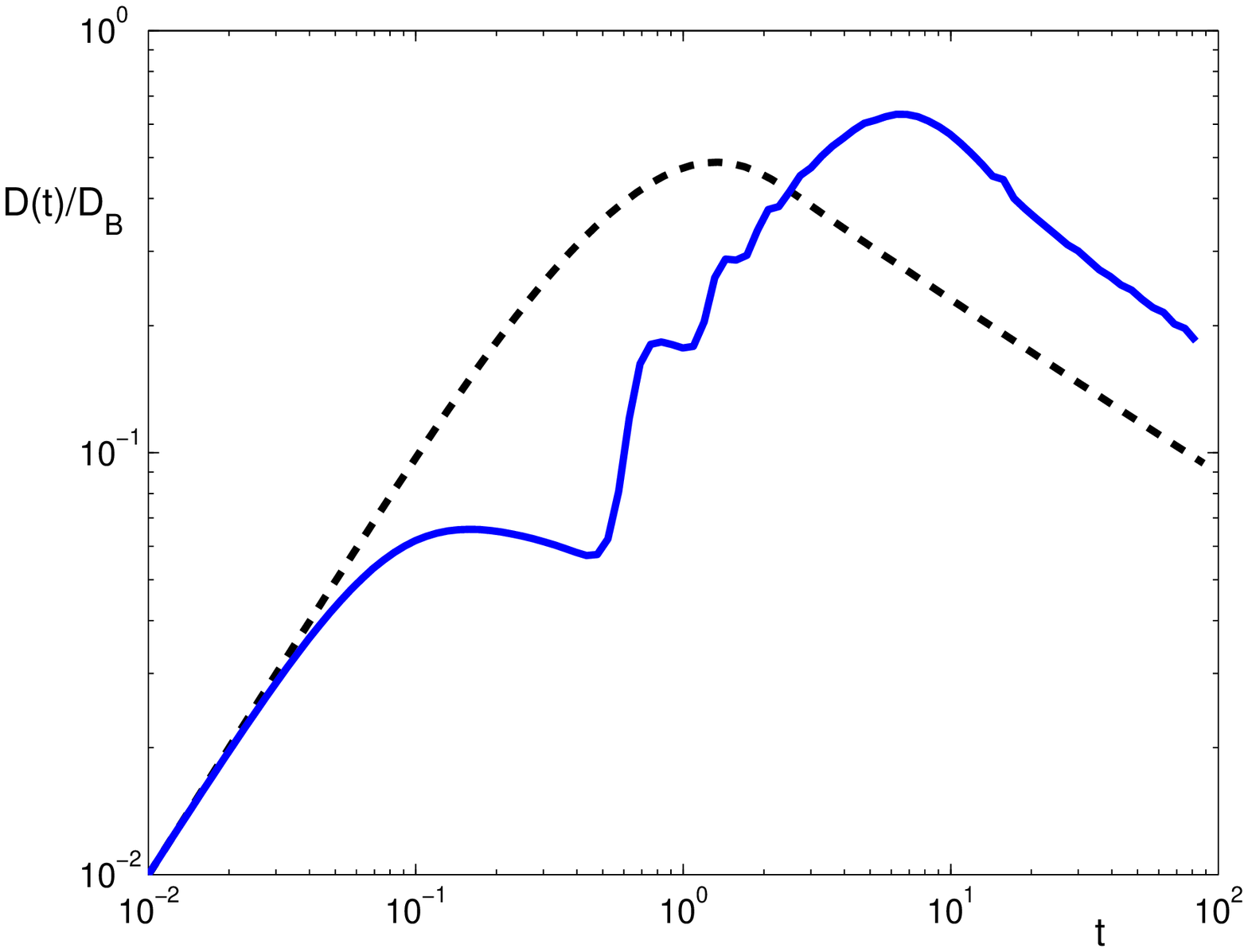}}


Figure 2: The time dependent diffusion coefficient for the Lorentz transport
(blue line) and for the drift transport (dashed line) for $\overline{%
\Omega }=10,$ $\overline{\rho }=1,$ $K=\infty .$
\end{center}

\vspace*{0.1in}

The correlation of the Lagrangian drift velocity is presented in Fig. 3 for
the Lorentz transport (continuous line) compared with drift approximation
(dashed line). It decays very fast (in a time much smaller than $\tau _{fl})$
and then it presents a series of peaks with decreasing amplitude and
eventually has a negative tail. The peaks appear around multiples of the
cyclotron gyration period $T=2\pi /\overline{\Omega }.$

\vspace*{0.1in}

\begin{center}
\resizebox{3.8in}{!}{\includegraphics{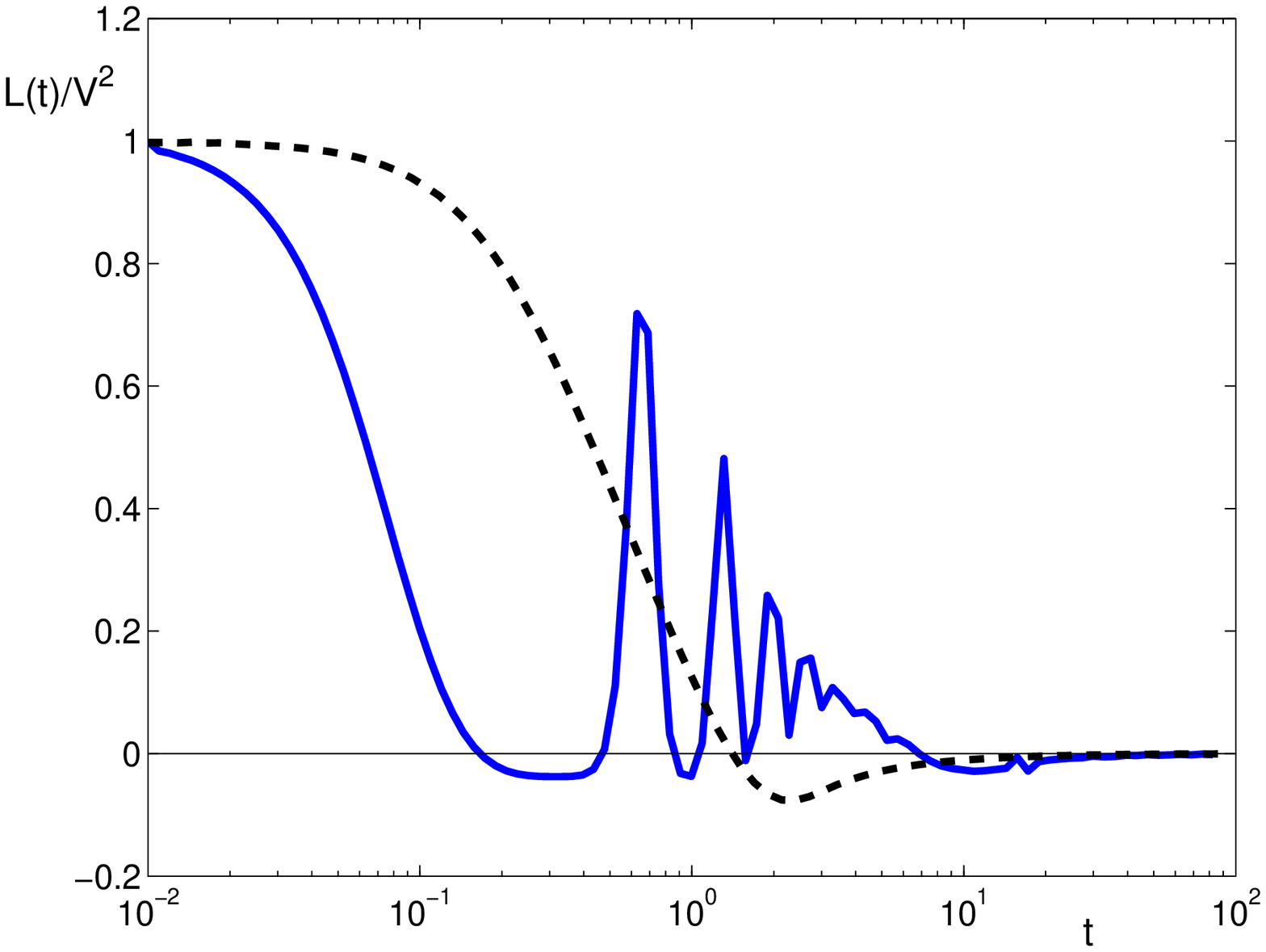}}


Figure 3: The correlation of the Lagrangian drift velocity for the Lorentz
transport (blue line) and for the drift transport (dashed line) for $%
\overline{\Omega }=10,$ $\overline{\rho }=1,$ $K=\infty .$
\end{center}

\vspace*{0.1in}

A clear story of the physical process can be deduced from the time evolution
of $D(t)$ and $L(t).$ Starting from $t=0$, at very small time ($t\ll T)$ $%
D(t)$ is equal with the drift diffusion coefficient. Then, the cyclotron
motion with a large Larmor radius ( $\overline{\rho }=1$ in Figs. 2, 3)
averages the stochastic field along the trajectory. Consequently the guiding
center has a very small displacement and $D(t)$ is much reduced compared to
the drift case. After a period the trajectories come back near the initial
position (all in phase because $T$ is a constant), a coherent motion of the
guiding centers appears during the passage of the particles and a step in $%
D(t)$ is produced. Thus the evolution of the guiding centers is determined
mainly by short coherent kicks appearing with period $T.$ Their displacement
is thus slower. Consequently, the trapping appears at later time and the
linear increase of $D(t)$ extends to longer times leading to values of $D(t)$
that are higher than for the drift diffusion. When the displacements of the
guiding centers increase the coherence of the periodic kicks is
progressively lost and the peaks of $L(t)$ becomes smaller and thicker.

The diffusion coefficient obtained for the Lorentz transport is a function
of the dimensionless parameters $K,$ $\overline{\rho }$ and $\overline{%
\Omega }.$ They contain the physical parameters of the stochastic field
(amplitude, correlation length, correlation time) and of the impurity ions
(mass, charge, kinetic energy). The dependence of the diffusion coefficient
on these parameters is analyzed in the next Sections.

\section{$K$ dependence}

We consider here time dependent stochastic potentials and determine the
dependence of the diffusion coefficient on the parameter $K$ defined in Eq. (%
\ref{K}).

In the case of drift transport, a change of variable can be done in the
equation for the decorrelation trajectories in order to introduce the time
factor of the Eulerian correlation of the potential in the time variable.
The diffusion coefficient for the time dependent potential is so determined
from $D(t)$ obtained for the static potential with the same space dependence
in the Eulerian correlation \cite{PPCF04}. The equations for the
decorrelation trajectories (\ref{1csiS}-\ref{1roS}) obtained for the Lorentz
transport do not have this property: due to the cyclotron motion (second
term in Eq. (\ref{1roS})), the time factor in the average potential $\Phi $
cannot be introduced in the time variable. Thus the $K$ dependence of the
diffusion coefficient must be determined by performing the calculations of $%
D(t)$ for each value of $K.$

\vspace*{0.10in}

\begin{center}
\resizebox{3.2in}{!}{\includegraphics{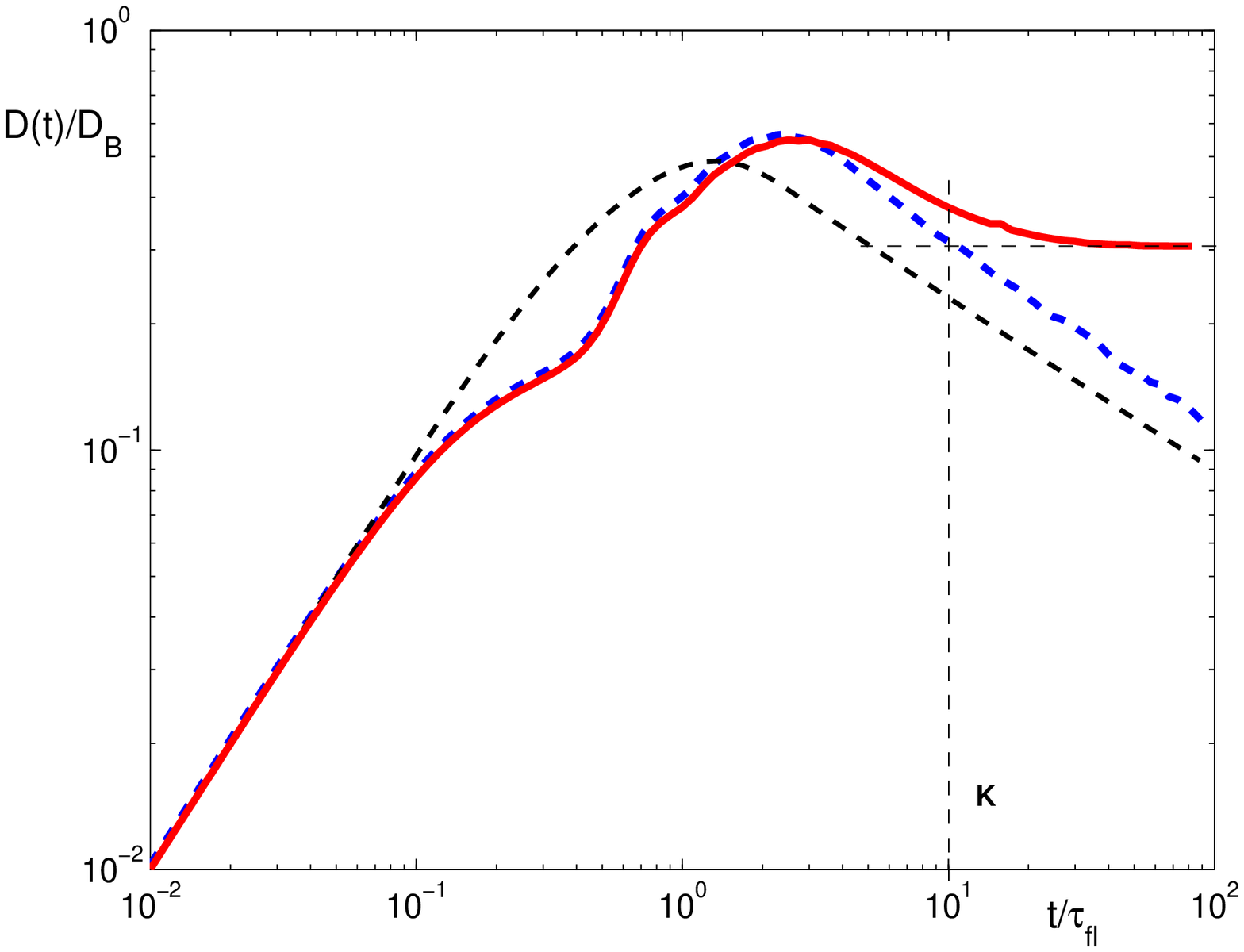}}


a
\end{center}


\begin{center}
\resizebox{3.2in}{!}{\includegraphics{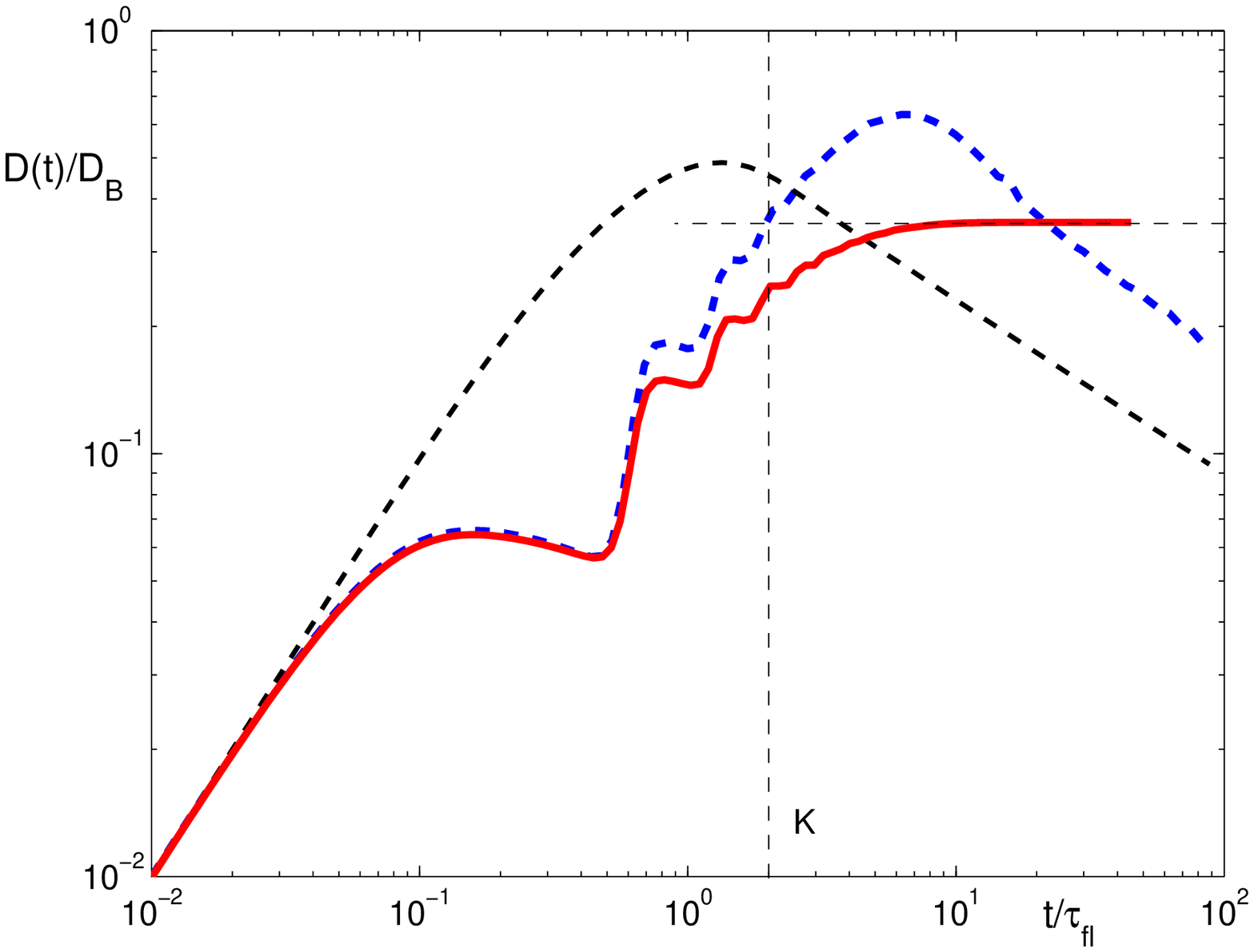}}


b

Figure 4: The evolution of the diffusion coefficient for the Lorentz
transport in a time dependent stochastic potentia (red line) compared
to the results obtained in a static potential for Lorentz transport (blue
line) and for the drift transport (dashed line).
\end{center}

However, we have shown in \cite{VS03}, using general considerations based on
the shape of the correlation of the Lagrangian velocity, that at large $K$
the asymptotic diffusion coefficient can be approximated by the diffusion
coefficient determined for the static potential at $t=\tau _{c},$ (or in
dimensionless units at $t=K).$ These considerations are not dependent on the
specific type of motion or on the statistical method used to obtain the
Lagrangian correlation. We examine here the accuracy of this approximation
for the Lorentz transport.

Typical examples for the evolution of the diffusion coefficient in time
dependent stochastic fields are presented in Fig. 4 (continuous line)
compared to $D(t)$ obtained in the static potential for Lorentz (dotted
line) and drift (dashed line) transport. One can see that the diffusion
coefficient saturates showing that the transport is diffusive in time
dependent stochastic potentials. A large $K$ case is considered in Fig. 4. a
which shows that the above approximation is rather accurate. For smaller $K$
values that are not situated on the tail of the Lagrangian correlation, the
demonstration presented in \cite{VS03} does not apply. However, as seen in
Fig. 4. b for $K=2$ the above approximation is valid and so was in many
other cases we have considered. Thus the asymptotic diffusion coefficient in
a time dependent stochastic potential with Kubo number $K$ is 
\begin{equation}
D(\infty |K,\overline{\rho },\overline{\Omega })\cong D(K|\infty ,\overline{%
\rho },\overline{\Omega })  \label{das}
\end{equation}
where $D(t|K,\overline{\rho },\overline{\Omega })$ is the time dependent
diffusion coefficient obtained for the parameters $K,\overline{\rho },%
\overline{\Omega },$ $D(\infty |K,\overline{\rho },\overline{\Omega })$ is
its asymptotic value and $D(t|\infty ,\overline{\rho },\overline{\Omega })$
is the diffusion coefficient obtained in the static potential.

\section{$\overline{\protect\rho }$ dependence}

The parameter $\overline{\rho }=\left| \mathbf{\rho }(0)\right| /\lambda
_{c} $ \ essentially describes the effect of the initial kinetic energy of
the ions on their Lorentz diffusion. It does not appear in the drift
transport, which is determined only by the Kubo number $K.$

The dependence of the asymptotic diffusion coefficient on the Kubo number
for several values of $\overline{\rho }$ is presented in Fig. 5. One can see
that the Larmor radius produces observable effects even for rather small
values (at $\overline{\rho }=0.1)$. As expected, the effect strongly
increase with the increase of $\overline{\rho }.$ This modification of the
diffusion coefficient due to Larmor radius is complex and it may consist of
a strong decrease as well as of a strong increase, depending on the
conditions. Thus, the general idea that the effective diffusion is reduced
due to the cyclotron motion which averages the stochastic potential, is not
always true.

\begin{center}
\resizebox{4.5in}{!}{\includegraphics{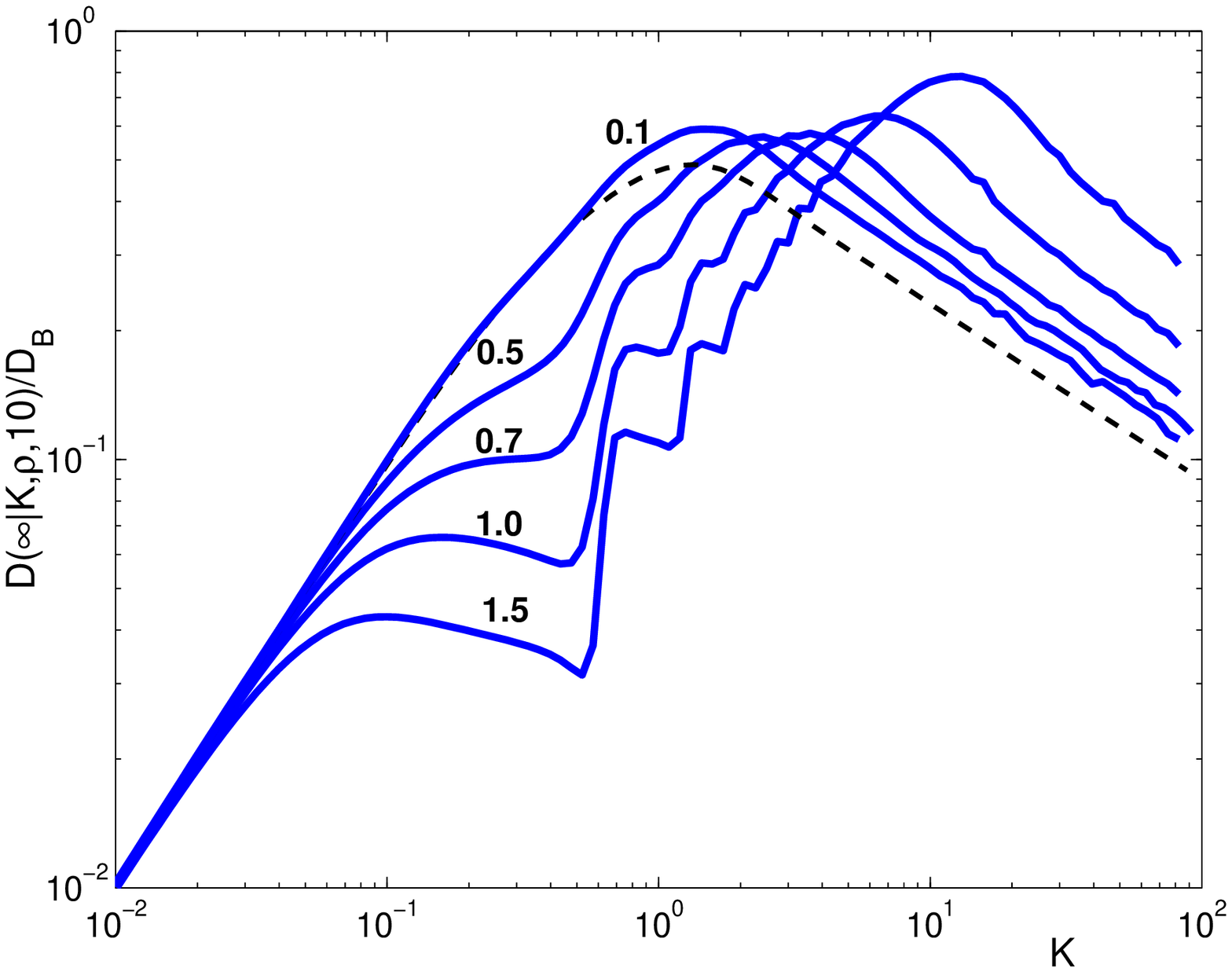}}


Figure 5: The asymptotic diffusion coefficient for the Lorentz transport as
a function of $K$ for several values of $\overline{\rho }$ that label the
curves and for $\overline{\Omega }=10.$ The result obtained in the guiding
center approximation is also represented (dashed line).
\end{center}

\bigskip

At small Kubo numbers the diffusion coefficient is much smaller than in the
drift approximation. It increases in steps that appear, for all values of $%
\overline{\rho },$ at values of $K$ which are multiples of the cyclotron
period $T=2\pi /\overline{\Omega }.$ Apart from these steps that are
attenuated at larger $K,$ there is a global increase with $K$ as $D(\infty
|K,\overline{\rho },\overline{\Omega })\sim D_{B}K=(\lambda _{c}^{2}/\tau
_{c})K^{2}$ in this regime. This is similar with the quasilinear regime of
the drift transport and corresponds to initial ballistic motion of the
guiding centers. This regime extends to  $K>1,$ up values that increase with
the increase of  $\overline{\rho }.$ The diffusion coefficient at this value
of $K$ has a value much larger than for the drift transport.

At larger values of \ $K$, the trapping becomes effective and the diffusion
coefficient has approximately the same $K$ dependence as in the drift
transport. The effect of the Larmor radius consists in an amplification
factor in the diffusion coefficient that is independent of $K$ in this
regime. It increases with the increase of $\overline{\rho }.$

\section{$\overline{\Omega }$ dependence}

The parameter $\overline{\Omega }=\Omega \tau _{fl}$ describes the effect of
the specific charge of the ions $q/m$ on their Lorentz diffusion. It
determines the moments of the steps appearing in the $K$ dependence of the
diffusion coefficient. Apart this, there is no strong influence of this
parameter when $\overline{\Omega }\gg 1$ (see Fig. 6). For small $\overline{%
\Omega }$ the trajectories become chaotic and the diffusion coefficient has
an irregular dependence on the parameters. Note that the chaotic variations
seen in this Figure at large $K\,$\ for $\overline{\Omega }=1$ are not
calculation errors (they are not changed by increasing the number of
calculated trajectories, thus the accuracy).

\begin{center}
\resizebox{3.9in}{!}{\includegraphics{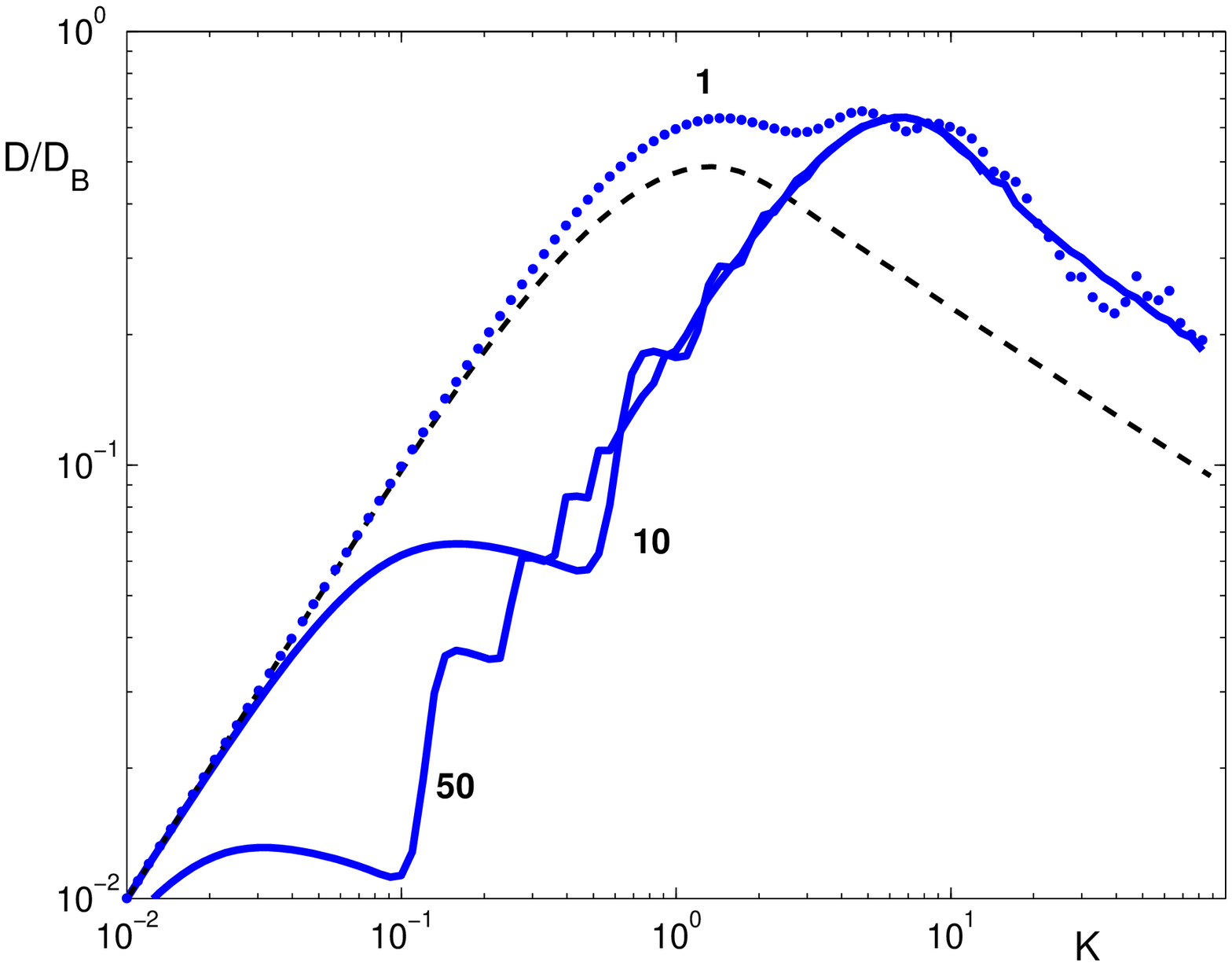}}


Figure 6: The asymptotic diffusion coefficient for the Lorentz transport as
a function of $K$ for the values of $\overline{\Omega }$ that label the
curves and for $\overline{\rho }=1;$ the result obtained in the guiding
center approximation is also represented (dashed line).
\end{center}

\section{Conclusions}

We have studied the impurity ion transport produced by the Lorentz force in
a turbulent magnetized plasma. Expressions for the time dependent diffusion
coefficient and for the correlation of the Lagrangian drift velocities are
obtained in terms of a class of smooth, deterministic trajectories by
developing a generalization of the decorrelation trajectory method. This
statistical approach is compatible with the invariance of particle energy.

We have shown that the Larmor radius has a strong effect on impurity ion
transport in turbulent plasmas. The generally accepted idea that the
effective diffusion is reduced due to the cyclotron motion which averages
the stochastic potential, is not always true. The cyclotron motion can also
determine the build up of correlation of the Lagrangian drift velocity by
bringing the particles back in the correlated zone of the stochastic
potential. The correlation $L(t)$ shows a series of periodic peaks, which
lead to incresed diffusion coefficients in slowly varying potentials.
Consequently, at given Larmor radius, the transport can be reduced or
increased, depending essentially on the value of the Kubo number.


\end{document}